\documentclass[11pt,reqno]{amsart}  
\allowdisplaybreaks[1]  

\usepackage{a4wide}  
\theoremstyle{plain} 
\newtheorem{theorem}{Theorem} 
 
\newtheorem{prop}{Proposition} 
\newtheorem{coro}{Corollary} 
 
\theoremstyle{definition} 
\newtheorem{definition}{Definition} 
 
\theoremstyle{remark} 
\newtheorem{remark}{Remark} 
 
\newcommand{\ts}{\hspace{0.5pt}} 
\newcommand{\PP}{\mathbb{P}} 
\newcommand{\RR}{\mathbb{R}\ts} 
 
\newcommand{\ZZ}{\mathbb{Z}} 
\newcommand{\NN}{\mathbb{N}}

\newcommand{\gL}{\varLambda} 
 
\DeclareMathOperator{\D}{dom} 
 
\DeclareMathOperator{\dist}{dist} 
 
\newcommand{\cals}{\mathcal{F}(\ZZ^d)} 
\newcommand{\calP}{\mathcal{P} (\ZZ^d)} 
\newcommand{\calb}{\mathcal{B} (\ZZ^d)} 
\newcommand{\calbnull}{\mathcal{B}_0 (\ZZ^d)} 
\newcommand{\cala}{\mathcal{A}} 
 
\newcommand{\calpnull}{\mathcal{P}_0^B } 
 
\newcommand{\calpM}{\mathcal{P}_0^B (M)} 
 
\newcommand{\calh}{\mathcal{H}} 
\newcommand{\calv}{\mathcal{V}} 
\newcommand{\lr}{\mathcal{L}({\RR})} 
\newcommand{\cD}{\mathcal{D}} 
\newcommand{\cF}{\mathcal{F}} 
\newcommand{\cH}{\mathcal{H}} 
\newcommand{\cL}{\mathcal{L}} 
\newcommand{\cM}{\mathcal{M}} 
 
\newcommand{\CM}{{C_M}} 
 
\newcommand{\iq}{I(Q,x,M)} 
 
\newcommand{\iqj}{I(Q_j,x,M)} 
\newcommand{\dq}{\partial(Q,x,M)} 
\newcommand{\sq}{S(Q,x,M)} 
\newcommand{\zxn}{\ZZ_x^M}

\newcommand{\elzh}{\ell^2 (\ZZ^d,\calh)}

\newcommand{\Hm}[1]{\leavevmode{\marginpar{\tiny% 
$\hbox to 0mm{\hspace*{-0.5mm}$\leftarrow$\hss}% 
\vcenter{\vrule depth 0.1mm height 0.1mm width \the\marginparwidth}% 
\hbox to 
0mm{\hss$\rightarrow$\hspace*{-0.5mm}}$\\\relax\raggedright #1}}} 
 
\begin{document}

\title[Uniform existence of the IDS]{Uniform existence of the 
  integrated density of states\\ for models on $\ZZ^d$}

\author{Daniel Lenz} 
\address{Fakult\"at f\"ur Mathematik, D-09107 Chemnitz, Germany  } 
\email{dlenz@mathematik.tu-chemnitz.de } 
\urladdr{http://www.tu-chemnitz.de/mathematik/analysis/dlenz} 
 
\author{Peter M\"uller}  
\address{Institut f\"ur Theoretische Physik, 
  Friedrich-Hund-Platz 1, D-37077 G\"ottingen, Germany} 
\email{peter.mueller@physik.uni-goe.de }  
\urladdr{http://www.theorie.physik.uni-goe.de/\raisebox{-4pt}{\hspace{-1.5pt}{\large\~{}}}mueller}

\author{Ivan Veseli\'{c}} 
\address{Emmy-Noether-Programme of the Deutsche 
  Forschungsgemeinschaft\vspace*{-0.3cm} }  
\address{\& Fakult\"at f\"ur Mathematik, TU Chemnitz, Germany   } 
%\email{ivan.veselic@mathematik.tu-chemnitz.de } 
\urladdr{http://www.tu-chemnitz.de/mathematik/schroedinger/members.php}

\begin{abstract} We provide an ergodic theorem for certain 
  Banach-space valued functions on structures over $\ZZ^d$, which 
  allow for existence of frequencies of finite patterns.  As an 
  application we obtain existence of the integrated density of states 
  for associated finite-range operators in the sense of convergence of 
  the distributions with respect to the supremum norm. These results 
  apply to various examples including periodic operators, percolation 
  models and nearest-neighbour hopping on the set of visible points.  Our 
  method gives explicit bounds on the speed of convergence in terms of 
  the speed of convergence of the underlying frequencies. It uses 
  neither von Neumann algebras nor a framework of random operators on 
  a probability space. 
\end{abstract} 
 
\maketitle

\section{Introduction} 
 
This paper deals with  existence of averages of Banach-space valued functions 
on subsets of $\ZZ^d$ and applications to finite-range operators.  
Existence of averages of Banach-space valued functions plays a role in 
both the study of thermodynamic formalism and the investigation of 
equivariant operators. For Banach-space valued functions on 
tiling-type structures exhibiting a special form of (dis)order known 
as aperiodic order, this has been investigated in \cite{GH,Len2,LS1}. 
Aperiodic order has attracted a lot of attention in recent years (see 
\cite{BM,Jan,Sen} for background and recent surveys).  In the 
mentioned works the assumption of aperiodic order implies uniform 
existence of frequencies of finite patterns, i.e.\ existence of 
frequencies along arbitrary van Hove sequences. This is equivalent to 
unique ergodicity of the underlying dynamical system. The key idea 
introduced in \cite{GH} is then to decompose the underlying structure 
into nice pieces and use existence of frequencies of these pieces in 
order to conclude the desired existence of the averages.  This type of 
program has been worked out in specific contexts in the above-cited 
literature. 
 
The assumption of uniform existence of frequencies is well adjusted to 
aperiodic order. However, it will not be met in a more random context. 
Thus, the question arises to what extent similar ergodic-type theorems 
can be proven in a more general framework.  It is quite clear that an 
answer to this problem is useful in various random and geometric 
contexts, see for instance \cite{DodziukLPSV-06}.  Our first result, 
Theorem~\ref{ergodictheorem}, answers this questions for models on 
$\ZZ^d$.  It turns out that an analogue of the previously given 
results is valid including explicit bounds on convergence under the 
sole assumption of existence of frequencies along a fixed van Hove 
sequence. In particular, we do not even need a context of dynamical 
systems. The proof uses ideas from \cite{Len2}, where one-dimensional 
uniquely ergodic subshifts are considered. 
 
While Theorem~\ref{ergodictheorem} may be of independent interest, in 
the present paper we use it to study uniform existence of the 
integrated density of states (IDS) for certain random operators. To do 
so, we follow the strategy of \cite{LS1} and reduce the proof of 
uniform convergence of the IDS to the validity of a Banach-space 
valued ergodic theorem.  This leads to our second result, 
Theorem~\ref{ids}, which ensures convergence of the normalised finite-volume 
eigenvalue counting functions to the IDS with respect to the supremum 
norm. Moreover, the explicit bounds on convergence in 
Theorem~\ref{ergodictheorem} yield explicit bounds for the speed of 
convergence for the IDS. 
 
As existence of frequencies is our only assumption, our approach is 
rather flexible and simple. In particular, we do not need von Neumann 
algebras or traces.  We illustrate this flexibility by applying our 
abstract results to three particular situations: ~(i)~~ periodic 
operators, ~(ii)~~ percolation models and ~(iii)~~ discrete Laplacians 
on the set of visible points. We emphasise that there is no good 
dynamical system available in the third situation.

\textit{Note added.} When we were finishing up this work we learned 
about the recent preprint \cite{Ele06} of Gabor Elek entitled 
``Aperiodic order, integrated density of states and the continuous 
algebras of John von Neumann'' containing related results. Using von 
Neumann algebras and a construction of Godearl, Elek can prove uniform 
convergence of the IDS for various models including percolation models 
and operators on self similar graphs.

\section{Basic notions} 
This paper is about functions and operators on colourings over 
$\ZZ^d$. We start by introducing the necessary background and notation.  
 
%\bigskip 

The set of all finite subsets of $\ZZ^d$ is denoted by $\cals$. We 
write $\sharp A$ for the cardinality of a general set $A$. For the 
special case of $Q\in \cals$, however, we use the notation $|Q|$ instead of 
$\sharp Q$. We will be 
particularly interested in boxes.  A subset $Q = [a_1,b_1]\times\cdots 
\times [a_d,b_d] \in\cals$ with $a_j,b_j\in \ZZ$, $a_j\leq b_j$, 
$j=1,\ldots,d$ is called a \emph{box}. The set of all boxes is denoted 
by $\calb$. A box $Q = [a_1,b_1]\times \cdots\times [a_d,b_d]$ with 
$(a_1,\ldots, a_d)=(0,\ldots,0)$ is said to \emph{lie at the origin}. 
The set of all boxes at the origin is denoted by $\calbnull$. 
Given $M\in \NN$, we write 
\begin{equation*} 
  \CM :=\{x\in \ZZ^d :  0 \leq x_j \leq M-1, j=1,\ldots, d\} 
\end{equation*} 
for the box with edges of length $M-1$ that is centred at the origin and 
$\CM(a) := a + \CM$ for $a\in \ZZ^d$.   
 
Let $\cala$ be a finite set. A map $\varLambda \colon \ZZ^d 
\longrightarrow \cala$ is called an \emph{$\cala$-colouring} of 
$\ZZ^d$. A map $ P \colon Q(P) \longrightarrow \cala$ with $Q(P)\in 
\cals$ is called an \emph{$\cala$-pattern}. The set $Q(P)$ is called 
the \emph{domain of $P$}. The set of all $\cala$-patterns is denoted 
by $\calP$.  When the set $\cala$ is understood from the context we 
will just speak about colourings and patterns.  For a colouring 
$\varLambda$ and a $Q\in \cals$ we define $\varLambda \cap Q$ by 
$\varLambda \cap Q \colon Q \longrightarrow \cala, y \mapsto 
\varLambda(y)$. Similarly, for a pattern $P$ and a box $Q$ with $Q 
\subset Q(P)$, we define $ P\cap Q \colon Q\longrightarrow \cala, y 
\mapsto P(y)$. For a pattern $P$ and $x\in \ZZ^d$ we define the 
shifted map $x + P$ by $ x + P : x + Q(P) \longrightarrow \cala, x + y 
\mapsto P(y)$. A pattern $P$ with $Q(P)\in \calb$ is called a 
\emph{box pattern}.  If, in addition, $Q(P)$ lies at the origin, then 
$P$ is said to be a \emph{box pattern at the origin}.  The set of all 
box patterns at the origin will be denoted by $\calpnull$, and the set 
of all box patterns at the origin with domain $\CM$ will be denoted by 
$\calpM$. 
% We denote by $\calp$, $\calpnull$ and $\calpM$ the set of  
%all box patterns, all box patterns at the origin and all  box 
%patterns at the origin with support $\CM$ respectively. 
For  $P\in \calpnull$  and $P'\in \calP$ arbitrary we define 
the number of occurrences of the box pattern $P$ in $P'$ by  
\begin{equation*}  
  \sharp_P P' := 
  \sharp\Bigl\{x\in Q(P') : x + Q(P) \subset Q(P'), P'\cap \bigl(x + 
  Q(P)\bigr)=  x + P \Bigr\}. 
\end{equation*} 
 
For an arbitrary $Q\in \cals$ and $S\in \NN$ we introduce the 
\emph{$S$-boundary} of $Q$ as 
\begin{equation*} 
  \partial^S Q :=\{x\in\ZZ^d \setminus Q: \dist(x,Q)\leq S \}\cup 
  \{x\in Q : \dist(x,\ZZ^d\setminus Q)\leq S \}.  
\end{equation*} 
A sequence $(Q_j)_{j\in\mathbb{N}} \subset\cals$ is called a \emph{van 
  Hove sequence} in $\ZZ^{d}$, if  
\begin{equation*} 
  \lim_{j  \to \infty} \frac{|\partial^S  Q_j|}{|Q_j|}=0 
\end{equation*} 
for every $S\in \NN$. Finally, let $\varLambda \colon \ZZ^d\longrightarrow 
\cala$ be a colouring, $P\in \calpnull$ a box pattern at the origin 
and $(Q_j)_{j\in\mathbb{N}}$ a van Hove sequence in $\ZZ^{d}$. If the limit 
\begin{equation*} 
  \nu_P := \lim_{j\to \infty} \frac{1}{|Q_j|} \sharp_P (\varLambda\cap Q_j) 
\end{equation*} 
exists, it is called the \emph{frequency of $P$ in $\varLambda$ along 
  $(Q_j)_{j\in\mathbb{N}}$}.  Existence of frequencies of box patterns is the 
\textbf{only} assumption we will pose on the colourings we consider. 
This will be sufficient to derive an ergodic-type theorem for certain 
Banach-space valued functions, which are introduced in 
Definition~\ref{function} below.  
 
\begin{definition} 
  \label{boundaryterm}  
  A map $b \colon \cals \longrightarrow 
  [0,\infty)$ is called a \emph{boundary term} if $b(Q) = b( t+ Q)$ 
  for all $t\in \ZZ^d$ and $Q\in \cals$, $\lim_{j\to \infty} 
  |Q_j|^{-1} b (Q_j) =0$ for any van Hove sequence $(Q_j)$, and there 
  exists $D>0$ with $b(Q) \leq D |Q|$ for all $Q\in \cals$. 
\end{definition} 
 
\begin{definition} 
  \label{function} 
  Let $(X,\|\cdot\|)$ be a Banach space and $F \colon 
  \cals\longrightarrow X$ be given. 
 
  (a) The function $F$ is said to be \emph{almost-additive} if there 
  exists a boundary term $b$ such that 
  \begin{equation*} 
    \| F ( \cup_{k=1}^m Q_k ) - \sum_{k=1}^m F(Q_k)\| \leq \sum_{k=1}^m b 
    (Q_k), 
  \end{equation*} 
  for all $m\in\mathbb{N}$ and all pairwise disjoint sets $Q_k\in\cals$, 
  $k=1,\ldots,m$.  
 
  (b) Let $\varLambda :\ZZ^d \longrightarrow \cala$ be a colouring. 
  The function $F$ is said to be \emph{$\varLambda$-invariant} if 
  \begin{equation*}  
    F( Q) = F (t + Q) 
  \end{equation*} 
  for all $t\in\ZZ^{d}$ and all $Q\in\cals$ obeying $t+ (\varLambda 
  \cap Q) = \varLambda \cap (t + Q)$. In this case there exists a 
  function $\widetilde{F} :\calpnull \longrightarrow X$ on the set of 
  box patterns at the origin such that 
  \begin{equation*}  
    F( t+ Q) = \widetilde{F} \Bigl(-t + \bigl(\varLambda\cap (t + 
    Q)\bigr)\Bigr)  
  \end{equation*} 
  for all $t\in \ZZ^d$ and all $Q\in \calbnull$.  
 
  (c) The function $F$ is said to be \emph{bounded} if there exists a 
  finite constant $C>0$ such that 
  \begin{equation*}  
    \|F(Q)\| \leq C |Q| 
  \end{equation*} 
  for all $Q\in \cals$.  
\end{definition}

We will use the Banach-space valued ergodic theorem in 
Sect.~\ref{sect:ergodic} below to study certain types of 
operators. These operators will be introduced now. 
 
Let $\calh$ be a fixed Hilbert space with dimension $\dim (\calh)< 
\infty$ and norm $\|\cdot\|$. Then, 
\begin{equation*} 
  \elzh:=\{ u \colon \ZZ^d \longrightarrow \calh : \sum_{x\in \ZZ^d} 
  \|u(x)\|^2<\infty\} 
\end{equation*} 
is a Hilbert space. The \emph{support} of $u\in \elzh$ is the set 
of $x\in\ZZ^d$ with $u(x) \neq 0$. 
 
For $x\in \ZZ^d$, we define the natural projection $p_x :\elzh 
\longrightarrow \calh$, $u \mapsto p_x (u) := u(x)$. Let $i_x \colon 
\calh \longrightarrow \elzh$ be the adjoint of $p_{x}$. Similarly, for 
a subset $Q\subset \ZZ^d$ we define $\ell^2 (Q,\calh)$ to be 
the subspace of $\elzh$ consisting of elements supported in $Q$.  The 
projection of $ \elzh$ on $\ell^2 (Q,\calh)$ is denoted by 
$p_Q$ and its adjoint by $i_Q$. 
 
The operators and functions we are interested in are specified in the 
next two definitions.

\begin{definition}\label{operator}  
  Let $\cala$ be a finite set, $\varLambda\colon\ZZ^d\longrightarrow 
  \cala$ a colouring and $H \colon \elzh \longrightarrow\elzh$ a 
  selfadjoint operator. 
 
  (a) The operator $H$ is said to be of \emph{finite range}, if there 
  exists a length $R_{\mathit{fr}}>0$ such that $p_y H i_x =0$, 
  whenever $x,y\in \ZZ^d$ have distance bigger than $R_{\mathit{fr}}$. 
 
  (b) The operator $H$ is said to be \emph{$\varLambda$-invariant} if 
  there exists a length $R_{\mathit{inv}}\in \NN$ such that $p_y H i_x 
  = p_{y+t} H i_{x +t}$ for all $x,y,t\in \ZZ^d$ obeying 
  \begin{equation*} 
    t + \Bigl(\varLambda \cap \bigl(C_{R_{\mathit{inv}}} (x)\cup 
    C_{R_{\mathit{inv}}} (y)\bigr) \Bigr)    
    = \varLambda\cap \bigl (C_{R_{\mathit{inv}}} (x+t)\cup 
    C_{R_{\mathit{inv}}} (y+t)\bigr).  
  \end{equation*} 
 
  (c) If $H$ is both of finite range and $\varLambda$-invariant, the 
  number $R(H) :=\max\{ R_{\mathit{fr}}, R_{\mathit{inv}}\}$ is called 
  the \emph{overall range} of $H$. 
\end{definition} 
 
A given finite-range, $\varLambda$-invariant operator $H$ is fully determined 
by specifying finitely many $\dim(\calh) \times \dim(\calh)$ matrices $p_y H 
i_x$. In particular, such operators $H$ are bounded.  
 
\begin{definition}  
  Let $\lr$ be the Banach space of right-continuous, bounded functions 
  equipped with the supremum norm.  For a selfadjoint operator $A$ on 
  a finite-dimensional Hilbert space we define its \emph{cumulative 
    eigenvalue counting function} $n(A) \in 
  \lr$ by setting  
  \begin{equation*} 
    n(A) (\lambda):=\sharp\{\text{eigenvalues of $A$ not exceeding 
    $\lambda$}\} 
  \end{equation*} 
  for all $\lambda\in\mathbb{R}$, where each eigenvalue is counted 
  according to its multiplicity. 
\end{definition} 
 
We will be particularly concerned with functions of the form $n (p_Q H i_Q)$  
for finite-range operators $H$ and $Q\in \cals$.

\section{An ergodic theorem} 
\label{sect:ergodic} 
 
In this section, we present an ergodic theorem for certain Banach-space 
valued functions. It is the main abstract input upon which we base our 
results on uniform convergence of the integrated density of states in 
Sect.~\ref{sect:uniex}. We will consider the following situation: 
 
\begin{itemize} 
\item[(E)] Let $\cala$ be a finite set, $\varLambda \colon 
  \ZZ^{d}\longrightarrow \cala$ an $\cala$-colouring and 
  $(X,\|\cdot\|)$ a Banach-space. Let $(Q_j)_{j\in\mathbb{N}}$ be a van Hove 
  sequence such that for every $P\in \calpnull$ the frequency $\nu_P = 
  \lim_{j\to \infty} |Q_j|^{-1} \sharp_P (\varLambda \cap Q_{j})$ 
  exists. 
\end{itemize}

\begin{theorem} \label{ergodictheorem}  
  Assume (E). Let $F : \cals 
  \longrightarrow X$ be a $\varLambda$-invariant, almost-additive 
  bounded function. Then, the limits 
  \begin{equation*} 
    \overline{F} := \lim_{j\to\infty} \frac{F(Q_j)}{|Q_{j}|} = \lim_{M\to\infty} 
    \sum_{P\in \calpM} \nu_P \frac{\widetilde{F}(P)}{|C_M|}  
  \end{equation*} 
  exist and are equal. Moreover, for $j,M\in\NN$ the difference 
  \begin{equation*} 
    \Delta(j,M) :=\biggl\| \,\frac{F (Q_j)}{|Q_j|} \; -  \sum_{P\in\calpM} \nu_P 
      \frac{\widetilde{F} (P)}{|C_M|}\,\biggr\| 
  \end{equation*} 
  of the finite-volume approximants  of $\overline{F}$ satisfies the  estimate  
 \begin{equation} 
   \label{e-error} 
   \Delta(j,M) \leq  
   \frac{b(C_M)}{|C_M|}  +  (C +D) \frac{|\partial^M Q_j|}{|Q_j|} 
   + C \sum_{P\in\calpM} \biggl| \frac{\sharp_P\varLambda\cap Q_j}{|Q_j|} 
     - \nu_P\biggr|.   
 \end{equation} 
 Here, $D$ is given in Definition \ref{boundaryterm} and 
 $\widetilde{F}$, $b$ and $C$ are given in Definition \ref{function}. 
\end{theorem} 
 
In order to emphasise the explicit bounds on convergence we state the 
following corollary. It provides 
apriori bounds on speed of convergence.

\begin{coro}  
  \label{explicit} 
  Assume the situation of Theorem~\ref{ergodictheorem}. Then we have 
  \begin{equation*} 
    \biggl\|\overline{F} - \frac{1}{|Q_j|} F(Q_j)\biggr\|\leq 2 
    \frac{b(C_M)}{|C_M|} +  C \sum_{P\in\calpM} \biggl| \frac{\sharp_P 
        (Q_j\cap \varLambda)}{|Q_j|} - \nu_P\biggr| +  (C + 
    D) \frac{|\partial^M Q_j|}{|Q_j|}  
  \end{equation*}  
  for all $j,M\in\NN$ and  
  \begin{equation*}  
    \biggl\|\overline{F} -\sum_{P\in\calpM} \nu_P \frac{ \widetilde{F} 
        (P)}{|C_M|}\biggr\|\leq \frac{b(C_M)}{|C_M|}  
  \end{equation*}  
  for all $M\in \NN$. 
\end{coro}

\begin{proof}[Proof of Theorem \ref{ergodictheorem}]  
  We will derive the explicit bound \eqref{e-error} on $\Delta(j,M)$.  
  This bound gives immediately that  
  \begin{equation*} 
    \lim_{M\to \infty} \lim_{j\to\infty} \Delta(j,M) =0. 
  \end{equation*} 
  Due to  
  \begin{equation*}  
    \biggl\|\frac{F(Q_j)}{|Q_j|}  - \frac{F(Q_m)}{|Q_m|} \biggr\|\leq 
    \Delta(j,M) + \Delta(m,M) 
  \end{equation*}  
  for all $M\in\mathbb{N}$, this in turn shows that $ ( |Q_j|^{-1} 
  F(Q_j))_{j\in\mathbb{N}}$ is a Cauchy sequence and hence convergent.  This 
  convergence then implies convergence of $\sum_{P\in \calpM} \nu_P 
  \frac{ \widetilde{F} (P)}{|C_M|} $ to the same limit. 
 
  We are now going to prove \eqref{e-error}. By the triangle 
  inequality we have for arbitrary $j,M\in\NN$ that 
  \begin{equation}  
    \label{split} 
    \Delta(j,M) \leq D_1 (j,M) + D_2 (j,M) 
  \end{equation} 
  with  
  \begin{align} 
    D_1 (j,M) &:=\biggl\| \frac{F(Q_j)}{|Q_j|}  - \sum_{P\in\calpM} 
      \frac{\sharp_P  
        (\varLambda\cap Q_j)}{|Q_j|} \ 
      \frac{\widetilde{F}(P)}{|C_M|}\biggr\|, \intertext{and}  
    D_2 (j,M)&:=\sum_{P\in\calpM}  \ \biggl| \frac{\sharp_P 
        (\varLambda\cap Q_j)}{|Q_j|} - \nu_P\biggr| \ 
    \frac{\|\widetilde{F}(P)\|}{|C_M|}.  
  \end{align} 
  By the boundedness assumption on $F$ we have 
  $\frac{\|\widetilde{F}(P)\|}{|C_M|}\leq  
  C$ for any $P\in \calpM$.  Thus 
  \begin{equation} 
    \label{D2bound} 
    D_{2}(j,M) \le C \sum_{P\in\calpM}  \ \biggl| \frac{\sharp_P 
        (\varLambda\cap Q_j)}{|Q_j|} - \nu_P\biggr| 
  \end{equation} 
  provides the last term in the error estimate \eqref{e-error}. 
 
  In order to bound the contribution $D_1$, we first introduce 
  some notation.  For $M\in \NN$, a box $Q\in \calb$ and $x\in \ZZ^d$ 
  we consider a (disjoint) covering of $Q$ with boxes $C_{M}(a)$ 
  shifted from the origin by vectors $a$ in the scaled lattice 
  $\zxn:=x + (M\ZZ)^d$. We write 
  \begin{equation*} 
    \sq :=\{a\in \zxn : C_M(a) \cap Q \neq \emptyset\} 
  \end{equation*} 
  for the minimal set of shift vectors needed to generate such a  
  covering of $Q$. The set $\sq$ decomposes into two \emph{disjoint} 
  parts, 
  \begin{equation}  
    \label{union} 
    \sq = \dq \cup \iq, 
  \end{equation} 
  where  
  \begin{align*} 
    \iq &:=\{a\in \zxn : C_M(a)  \subset Q\}, \\ 
    \dq &:=\{a\in \zxn : a \notin \iq \;\:\mbox{and}\;\: C_M (a) \cap Q \neq 
    \emptyset\}.  
  \end{align*}   
  The interpretation of the sets $\iq$ and $\dq$ is as follows: the 
  boxes of the covering that are shifted by $a\in\iq$ are fully 
  contained in $Q$, while those shifted by $a\in\dq$ are only partly 
  contained in $Q$.  For later purpose we remark that 
  \begin{equation} 
    \label{boundcount} 
    |\dq| \le \frac{\partial^{M}Q}{|C_{M}|}. 
  \end{equation} 
  Given $M\in \NN$, $Q\in \calb$ and $x\in \ZZ^d$ 
  arbitrary, we then estimate 
  \begin{align}  
    \label{prelim} 
    T(Q,x,M) &:= 
    \biggl\| F(Q) - \sum_{a\in \iq} F\bigl(C_M (a)\bigr)\biggr\| \nonumber\\ 
    &\phantom{:} \leq  \biggl\|F(Q) - \sum_{a\in \sq} F\bigl(C_M(a) 
      \cap Q\bigr) \biggr\| + \biggl\|\sum_{a\in \dq} F\bigl(C_M (a) 
      \cap  Q\bigr)\biggr\| \nonumber\\   
    &\phantom{:}\leq    
     \left(\sum_{a\in \iq} b\bigl(C_M (a)\bigr) + 
      \sum_{a\in \dq} b\bigl(C_M (a) \cap Q\bigr)\right)    
    +   C \sum_{a\in \dq}  |C_M (a)| \nonumber\\ 
    &\phantom{:}\leq  |Q| \frac{b(C_M)}{|C_M|} +  |\dq| (C + D) |C_M| .  
  \end{align} 
  The first inequality in \eqref{prelim} follows from \eqref{union}, the 
  second from the almost-additivity of $F$ and the third from the 
  boundedness of $F$ and of the boundary term. 
 
  Now, we come back to $D_1$ defined in \eqref{D1bound}. From 
  \begin{equation*}  
    \sum_{P\in\calpM} \sharp_P (\varLambda\cap Q_j)  \ \widetilde{F}(P) = 
    \sum_{z\in \ZZ^d : C_M (z) \subset Q_j} F\bigl(C_M (z)\bigr)= 
    \sum_{x\in C_M} \sum_{a\in \iqj} F\bigl( C_M (a)\bigr) 
  \end{equation*} 
  we deduce 
  \begin{equation*} 
    \label{D1bound} 
    |Q_j| \, D_1 (j,M) = 
     \biggl\| F(Q_j) - \frac{1}{|C_M|} \sum_{x\in C_M} 
      \sum_{a\in \iqj} F\bigl( C_M (a)\bigr)\biggr\|   
    \leq \frac{1}{|C_M|} \sum_{x\in C_M} T(Q_j,x,M)  . 
  \end{equation*} 
  Therefore,  \eqref{boundcount} and  \eqref{prelim} give 
  \begin{equation*}  
    D_1 (j,M) \leq  \frac{b(C_M)}{|C_M|}   +  \frac{C + D}{|Q_{j}|} 
     |\partial^{M} Q_{j}| .  
  \end{equation*}  
  Together with \eqref{split} and \eqref{D2bound}, this finishes the 
  proof of \eqref{e-error}.  
\end{proof}

\begin{proof}[Proof of Corollary \ref{explicit}] 
  For every $j,M\in\mathbb{N}$ we have 
  \begin{equation*}  
    \biggl\|\overline{F} - \frac{F(Q_j)}{|Q_j|}  \biggr\| = 
    \lim_{m\to \infty} \biggl\| \frac{F(Q_m)}{|Q_m|}  - \frac{F(Q_j)}{|Q_j|} 
     \biggr\| \leq \limsup_{m\to \infty}  \bigl[ \Delta(m,M) + 
     \Delta(j,M) \bigr], 
  \end{equation*} 
  and the first estimate follows follows from 
  \eqref{e-error}. Similarly, we have   
  \begin{equation*}  
    \biggl\|\overline{F} - \sum_{P\in\calpM} \nu_P \frac{ 
      \widetilde{F} (P)}{|C_M|} \biggr\|   
    = \lim_{j\to\infty} \biggl\| \frac{F(Q_j)}{|Q_j|} 
       - \sum_{P\in\calpM} \nu_P   \frac{ \widetilde{F} 
        (P)}{|C_M|}\biggr\|   
    = \lim_{j\to\infty} \Delta(j,M) 
  \end{equation*} 
  for all $M\in\mathbb{N}$, and \eqref{e-error} implies the second 
  estimate.   
\end{proof}

\section{Uniform existence of the integrated density of states --  
  abstract results}  
\label{sect:uniex} 
 
In this section we assume the following situation:  
 
\begin{itemize} 
\item[(S)]  
  Let $\varLambda\colon\ZZ^d\longrightarrow \cala$ be a 
  colouring and $(Q_j)_{j\in\mathbb{N}}$ a van Hove sequence along which the 
  frequencies $\nu_P$ of all patterns $P\in \calpnull$ exist.  Let 
  $H:\elzh\longrightarrow\elzh$ be a selfadjoint, 
  $\varLambda$-invariant finite-range operator.  Let $R = R(H)$ denote 
  the overall range of $H$ as given in Definition \ref{operator}. 
\end{itemize} 
 
For certain statements it will be necessary to assume as well Condition 
\begin{itemize} 
\item[(+)]  
  The frequencies $\nu_P$ are strictly positive for all 
  patterns $P$ which occur in $\varLambda$, i.e.\ for which there 
  exists $x\in\ZZ^d$ with $\varLambda \cap (x + Q(P)) = x + P$. 
\end{itemize}

\begin{theorem}\label{ids}  
  Assume (S). Then, there exists a unique probability measure $\mu_H$ 
  on $\RR$ with distribution function $N_H$ such that $\frac{1}{|Q_j|} 
  n( p_{Q_j} H i_{Q_j})$ converges to $N_H$ with respect to the 
  supremum norm as $j\to\infty$. The distribution function $N_{H}$ is 
  called the \emph{integrated density of states (IDS)} of $H$. In 
  fact, the estimate 
  \begin{align} 
    \label{N-error1} 
    \left \| \frac{  n( p_{Q_j} H  i_{Q_j})  }{|Q_j|} - N_H \right\|_\infty  
    & \leq  8 \frac{|\partial^R C_M|}{|C_M|} \dim (\calh) +    
    \sum_{P\in\calpM}  \left| \frac{\sharp_P (\varLambda\cap Q_j)}{|Q_j|} - 
      \nu_P\right| \nonumber\\ 
    & \quad + 5 |C_R| \frac{|\partial^M Q_j|}{|Q_j|} \dim(\calh)  
  \end{align} 
  holds for all $j,M\in\NN$.   
  If $(+)$ holds as well, then the spectrum of $H$ 
  is the topological support of $\mu_H$. 
\end{theorem}

\begin{remark}  
  Assumption $(+)$ is necessary to obtain equality of the spectrum and 
  of the topological support of $\mu_H$. This can easily be seen from 
  examples. Take, e.g., the identity operator $\mathit{id}$ on 
  $\ell^2(\ZZ)$ and perform a rank one perturbation $B = \langle 
  \delta_0, \cdot\rangle \delta_0$ at the origin. Then, the IDS of 
  $\mathit{id}$ and of $\mathit{id} + B$ coincide, but their spectra 
  do not. 
\end{remark}

We also obtain an analogue of Corollary \ref{explicit} for the 
IDS. For this purpose, and for later use as well, we introduce 
for a given $S\in\NN$ the interior core 
\begin{equation*}  
  Q_S :=\{x\in Q : \dist(x,\ZZ^d \setminus Q)>S\} = Q \setminus 
  \partial^{S}Q 
\end{equation*} 
of a bounded set $Q\in\cals$. 
 
\begin{coro}\label{huhu}  
  Assume (S) and let $N_{H}$ be given by Theorem~\ref{ids}. For 
  $M\in \NN$ and a pattern $P\in \calpM$ with $\nu_P>0$ choose $x\in 
  \ZZ^d$ such that $x + P = \varLambda \cap \bigl(x+ Q(P)\bigr)$ and 
  define $n_P\in  
  \lr$ by 
  \begin{equation*} 
    n_P := n \big( p_{x+(Q(P))_R}  H \, i_{x+(Q(P))_R}\big). 
  \end{equation*}  
  Then, $n_P$ does not depend on the choice of $x$ and the bound 
  \begin{equation}  
    \label{N-error2} 
    \biggl\| N_H - \sum_{P\in \calpM} \nu_P \, \frac{n_P}{|C_M|} \biggr\|_{\infty} \leq 8 
    \dim(\calh) \,   \frac{|\partial^R C_M|}{|C_M|}  
  \end{equation} 
  holds for all $M\in \NN$.    
\end{coro}

\begin{remark} 
  We emphasise that the dependence on $H$ of the error bounds 
  \eqref{N-error1} and \eqref{N-error2} is very weak. In fact, it is 
  only the overall range $R$ of $H$ which enters. Thus, our 
  estimates work simultaneously for all $H$ with the same $R$. 
\end{remark}

\begin{coro} 
  \label{jump}  
  Assume (S) and (+), and let $\lambda\in \RR$. Then the following 
  assertions are equivalent: 
  \begin{itemize} 
  \item[(i)] $\lambda$ is a point of discontinuity of $N_H$, 
  \item[(ii)] there exists a compactly supported eigenfunction of $H$ 
  corresponding to $\lambda$.   
  \end{itemize} 
\end{coro}

The remainder of this section is devoted to a proof of these results. We begin 
with some preliminary considerations.

\begin{prop}[See Prop.~5.2 in \cite{LS1}] 
  \label{subspace}  
  Let $U$ be a subspace of a finite-dimensional Hilbert space $V$ with 
  inclusion $i \colon U \longrightarrow V$ and orthogonal projection 
  $p \colon V\longrightarrow U$. Then, 
 \begin{equation*} 
   \|n(A) - n(p A i)\|_{\infty} \leq  4\cdot \mathrm{rank} (1-i\circ p)  
 \end{equation*} 
 for every selfadjoint operator  $A$ on $V$.    
\end{prop}

\begin{prop} 
  \label{AAT}  
  Assume (S). Then $F_R^H \colon \cals \longrightarrow \lr$, $ Q \mapsto 
  F_R^H (Q) := n( p_{Q_R} H i_{Q_R} )$, is an almost-additive, 
  bounded and $\varLambda$-invariant function with $C=1$ and boundary 
  term $b (Q) := 4 |\partial^R Q| \dim(\calh)$. 
\end{prop}

\begin{proof} 
  This is a consequence of the previous proposition. Boundedness with 
  constant $C=1$ is clear from the definition of $n (\cdot)$. 
  Similarly, $\varLambda$-invariance follows by construction of $F^H_{R}$ 
  (namely, $F^H_{R} (Q)$ only depends on $Q \cap \varLambda$). 
  Almost-additivity follows from a decoupling argument (see \cite{LS1} 
  as well). More precisely, let $Q$ be given such that $Q$ is the 
  disjoint union of $Q_k$ for $k=1,\ldots,m$. By our choice of $R$ we 
  then have 
  \begin{equation*}  
    p_{ \cup_{k=1}^m Q_{k,R}} H  i_{\cup_{k=1}^m Q_{k,R}}  
    = \bigoplus_{k=1}^m  \ p_{Q_{k,R}} H i_{ Q_{k,R} }  
  \end{equation*} 
  and hence 
  \begin{equation*}  
    n(  p_{\cup_{k=1}^m Q_{k,R}} H  i_{\cup_{k=1}^m Q_{k,R}})  
    =  \sum_{k=1}^m n( p_{Q_{k,R}} H i_{Q_{k,R}}).  
  \end{equation*} 
  Moreover,  Proposition \ref{subspace}  shows that  
  \begin{equation*}  
    \|n(p_{Q,R} H i_{Q,R})  
    - n( p_{\cup_{k=1}^m Q_{k,R}} H  i_{\cup_{k=1}^m  Q_{k,R}})\|_\infty  
    \leq  4 \sum_{k=1}^m |\partial^R Q_k| \dim (\calh). 
  \end{equation*} 
  This gives almost-additivity of $F^H_{R}$ with  the boundary term 
  \begin{equation*}  
    b (Q) :=  4 |\partial^R Q| \dim(\calh), 
  \end{equation*} 
  and the proof is complete.  
\end{proof}

\begin{prop} 
  \label{multiplicity}  
  Let $A$ be a selfadjoint operator on a finite-dimensional Hilbert space $V$. 
  Let $\lambda \in \RR$ and $\varepsilon>0$ be given and denote by $U$ the 
  subspace of $V$ spanned by the eigenvectors of $A$ belonging to eigenvalues 
  in the open interval $(\lambda-\varepsilon,\lambda +\varepsilon)$.  If there 
  exist $k$ pairwise orthogonal and normalised vectors $u_1,\ldots,u_k \in V$ 
  such that $(A- \lambda) u_j$, $j=1,\ldots, k$, are pairwise orthogonal and 
  satisfy $\|(A -\lambda) u_j\| < \varepsilon$, then $\dim(U) \ge k$. 
\end{prop}

\begin{proof}  
  Denote the linear span of $u_1,\ldots, u_k$, by $S$, and let $P \colon S 
  \longrightarrow U$ be the orthogonal projection from $S$ to $U$.  It suffices 
  to show that $P$ is one-to-one. Assume the contrary, then there exists a unit 
  vector $u\in S$ which is orthogonal to $U$.  This orthogonality yields $\| 
  (A - \lambda) u\|\geq \varepsilon$.  On the other hand, the assumptions on 
  the $u_j$, $j=1,\ldots, k$ imply $\| (A - \lambda) u\|< \varepsilon$. 
\end{proof}

Now we are prepared for the

\begin{proof}[Proof of Theorem \ref{ids}]  
  The statement on convergence is a direct consequence of Proposition 
  \ref{AAT} and Theorem \ref{ergodictheorem}. The explicit bound then 
  follows from Corollary \ref{explicit} and Proposition \ref{AAT}. 
   
  The last statement on the topological support follows by 
  Weyl-sequence-type arguments. Here are the details: first, note that   
  \begin{equation} 
    \label{starstar} 
    \|( p_{Q} H i_{Q } -\lambda)u )\| =\|(H - \lambda) u\| 
  \end{equation} 
  whenever $Q$ is a finite subset of $\ZZ^d$ and $u$ is supported in 
  $Q_R$, as $H$ is of range $R$. 
 
  Let now $\lambda$ belong to the spectrum $\sigma(H)$. Then, for each 
  $\varepsilon>0$ we can find a box $Q$ and a normalised vector $u$ 
  with support in $Q_{R}$ and $\|(H- \lambda) u\|< \varepsilon$. As 
  $H$ has range $R$ we infer that both $u$ and $(H-\lambda) u$ are 
  supported in $Q$ and $(p_Q H i_Q -\lambda) u$ has norm strictly less 
  than $\varepsilon$ by \eqref{starstar}.  Assume now that there are 
  $k$ disjoint occurrences of translates of $\varLambda \cap Q$ in a 
  set $Q_j$ for $j\in \NN$. These will provide mutually orthogonal 
  normalised functions $u_j$, $j=1,\ldots,k$, with $(H- \lambda)u_j$ 
  pairwise orthogonal and of norm strictly less than $\varepsilon$. 
  Proposition \ref{multiplicity} shows 
  \begin{equation} 
    n(p_{Q_j} H i_{Q_j})(\lambda+ \varepsilon) - n(p_{Q_j} H 
    i_{Q_j})(\lambda - \varepsilon)\geq k. 
  \end{equation} 
  By the assumption of positivity of the frequencies we see that the 
  number of disjoint occurrences of a certain pattern grows linearly 
  in the volume of $Q_j$ for large $n$, i.e. $k\geq c |Q_j|$ with 
  $c>0$ suitable. By uniform convergence of the $ n(p_{Q_j} H 
  i_{Q_j})$ this gives $\mu_H ([\lambda-\varepsilon, \lambda + 
  \varepsilon]) \geq c>0$. As $\varepsilon>0$ is arbitrary, we infer 
  that $\lambda$ belongs to the support of $\mu_H$. 
 
  Conversely, let $\lambda$ belong to the support of $\mu_H$. Then 
  there exists for each $\varepsilon >0$ a $c>0$ with $\mu_H 
  ([\lambda-\varepsilon, \lambda+ \varepsilon])\geq c$. By uniform 
  convergence this gives that 
  \begin{equation*} 
    n(p_{Q_j} H i_{Q_j}) ( \lambda+ \varepsilon) -  
    n(p_{Q_j} H i_{Q_j}) ( \lambda - \varepsilon)\geq \frac{c}{ 2} 
    |Q_j| 
  \end{equation*}  
  for sufficiently large $n$.  Thus, by standard linear algebra (see 
  \cite{KLS} for similar reasoning), for sufficiently large $n$ we can 
  find a normalised $u$ which is compactly supported in $Q_{n,R}$ and 
  satisfies 
  \begin{equation*}  
    \| (p_{Q_{n}} H i_{Q_{n}} -\lambda)u)\|\leq \varepsilon. 
  \end{equation*} 
  Then, $ \|(H - \lambda) u\| \leq \varepsilon $ by \eqref{starstar}, 
  and we see that $\sigma (H) \cap [\lambda -\varepsilon, \lambda + 
  \varepsilon] \neq \emptyset$. As $\varepsilon >0$ is arbitrary, we 
  infer that $\lambda$ belongs to $\sigma (H)$. 
\end{proof}

\begin{proof}[Proof of Corollary \ref{huhu}] This is a direct consequence of 
  Proposition \ref{AAT} and Corollary \ref{explicit}.  
 \end{proof}

\begin{proof}[Proof of Corollary \ref{jump}] 
 The corollary follows from Theorem~\ref{ids} by mimicking the argument 
  given in \cite{KLS} to prove Theorem $2$ there.  
\end{proof}

%%%%%%%%%%%%%%%%%%%%%%%%%%%%%%%%%%%%%%%%%%%%%%%%%%%%%%%%%% 
 
\section{Application to  periodic operators} 
\label{s-periodic} 
 
In this section we apply the above abstract results to periodic 
operators on graphs with a $\ZZ^d$-structure.  It turns out that for a 
large class of such operators, it is sufficient to consider the case 
where the set $\cala$ has just one element.  So we turn to this 
situation now. 
 
If $|\cala|=1$, the colouring $\varLambda$ is a trivial map,  
and the only information that is contained in an $\cala$-pattern $P$  
is its domain $Q(P)$, which is by definition a finite subset of $\ZZ^d$.   
For $M$ twice as large as the diameter of $Q(P)$ and 
for any sequence van Hove sequence $(Q_j)_{j\in\mathbb{N}}$ we have 
\begin{equation*} 
  1 \ge \frac{\sharp_P (\varLambda\cap Q_j)}{|Q_j|}   
  \ge\frac{|Q_{j,M}|}{|Q_j|}  \to 1 \quad \text{ for } j \to \infty . 
\end{equation*} 
Thus the frequency of any pattern $P$ equals $\nu_P=1$.  In this 
situation, where the frequencies of patterns happen to exist along any 
van Hove sequence, it is particularly convenient to chose them as 
boxes, more precisely $Q_j=C_j$ for all $j \in \NN$. 
 
Now, for a bounded, selfadjoint operator $H$ as in Definition 
\ref{operator}, Theorem \ref{ids} gives for all $j,M\in\NN$ the 
estimate 
\begin{equation*} 
  \left \| \frac{  n( p_{C_j} H  i_{C_j})  }{j^d} - N_H \right\|_\infty  
  \leq  8 \, d \, \dim(\calh) \ \Big (\frac{4R}{M} + \frac{5R^dM}{j} \Big) 
  + \frac{dM}{j}\,. 
\end{equation*} 
Similarly as in Corollary \ref{huhu} we set  
$n_{C_M}:= n\big( p_{C_{M,R}}  H \, i_{C_{M,R}}\big) \in\cL(\RR)$ 
and obtain thus  
\begin{equation*}  
  \| N_H -  \frac{n_{C_M}}{|C_M|} \|_{\infty} \leq  \dim(\calh) \ \frac{16 d R}{M} 
\end{equation*} 
for all $M\in \NN$. 
 
Now we describe the geometry of the class of graphs with 
$\ZZ^d$-structure on which we will define our periodic operators later 
on.  Let $G$ be a graph with a countable set of vertices (which we again 
denote by $G$), and $T$ a representation of $\ZZ^d$ by isometric 
graph-isomorphisms $T_\gamma\colon G\to G$, $\gamma \in \ZZ^d$.  We 
assume that the action $T$ of $\ZZ^d$ on $G$ is free and cocompact. 
 
Let us denote by $\mathcal{D}\subset G$ a %connected 
$\ZZ^d$-fundamental domain. Thus $\mathcal{D}$ contains exactly one 
element of each $\ZZ^d$-orbit in $G$.  By the cocompactness 
assumption, $\mathcal{D}$ is finite.  This implies in particular that 
the vertex degree of $G$ is uniformly bounded.  From now on the 
fundamental domain $\cD$ will be assumed fixed. 
 
A simple example of such a graph is $\ZZ^d$ with the group $\ZZ^d$ 
acting on it by $T_\gamma(x)=x- N\gamma$, $N\ge1$, for all $x \in 
\ZZ^d$ and all $\gamma \in \ZZ^d$.  Another example would be the 
Cayley graph $G=G(\mathcal{G},E)$ of a direct product group 
$\mathcal{G}= \ZZ^d\otimes F$ where $F$ is any finite group and $E$ is 
a symmetric set of generators for $\mathcal{G}$.  The action of 
$\ZZ^d$ on $G$ is given by 
\begin{equation*} 
  T_\gamma(x,f)=(x-\gamma,f), \quad (x,f) \in \mathcal{G}, \gamma \in\ZZ^d,  
\end{equation*} 
on the set of vertices $\mathcal{G}$ and analogously on the set of edges. 
 
Now we introduce operators acting on $\ell^2(G)$ and $\elzh$.  Let $A 
\colon \ell^2(G) \to\ell^2(G) $ be a selfadjoint linear operator 
satisfying the following equivariance condition 
\begin{equation*} 
  A(x,y)= A(T_\gamma  x,T_\gamma  y)  
  \quad\text{ for all }  x,y \in G, \gamma \in \ZZ^d. 
\end{equation*} 
Furthermore we assume that $A(x,y) \neq 0 $ implies that the 
graph-distance of $x$ and $y$ is smaller than $\rho \in \RR$.

Set $\cH := \ell^2(\cD)$, then $\dim(\calh)=|\cD|$, and define a 
unitary operator $U \colon \elzh \to\ell^2(G)$ in the following way: 
for a $\psi \in \elzh$ and $\gamma\in\ZZ^d$ write $\psi(\gamma) 
=\sum_{i\in \cD} \psi_i(\gamma) \delta_i$ where $(\delta_i)_{i\in\cD}$ 
is the standard orthonormal basis of $\ell^2(\cD)$. Note that the 
coefficients $\psi_i(\gamma)$ are uniquely defined. Define now 
$U\psi\in \ell^2(G)$ by $(U\psi)(x):= \psi_i(\gamma)$ where $i\in \cD$ 
and $\gamma \in \ZZ^d$ are the unique elements such that $x= T_\gamma 
i$. The inverse $U^*$ is given by $(U^*\phi) (\gamma) = \sum_{i \in 
  \cD} \phi(T_\gamma i)\delta_i$ for $\phi\in \ell^2(G)$. 
 
Next we define the operator $H \colon \elzh \to\elzh$ by $H:= U^*AU$ 
and show that it is a $\varLambda$-invariant operator of finite range. 
Let $\alpha,\beta \in \ZZ^d$ be arbitrary.  Then $U i_\beta$ maps 
$\cH$ to $\ell^2(T_\beta \cD)$ and $p_\alpha U^*$ maps 
$\ell^2(T_\alpha \cD)$ to $\cH$. Since $G$ and its (abelian) covering 
transformation group $\ZZ^d$ are quasi-isometric, it follows that if 
we choose $R_{\mathit{fr}}>0$ sufficiently large, we have $|x-y 
|>\rho$ for all $x \in T_\alpha \cD, y \in T_\beta \cD, |\alpha-\beta 
|>R_{\mathit{fr}}$.  Thus $p_\alpha U^*AU i_\beta=0$ for 
$|\alpha-\beta |>R_{\mathit{fr}}$.  A straightforward calculation 
shows that $ p_{\alpha +\gamma} H i_{\beta +\gamma} =p_\alpha H 
i_\beta $.

%%%%%%%%%%%%%%%%%%%%%%%%%%%%%%%%%%%%%%%%%%%%%%%%%%%%%%%%%%%%%%%%%%%%%%%%%%%%%%%%%%%%%%%%%%%%%%%%% 
 
\section{Application to Anderson-percolation models} 
 
In this section we discuss similar operators as in the previous one,  
however now some randomness enters the model. 
In particular the colouring $\varLambda$ is no longer trivial.  
 
Let $G$ denote the graph $\ZZ^d\subset\RR^d$ where two vertices are 
adjacent, if and only if their Euclidean distance is equal to one. For some $N 
\in\NN$ fixed consider the action of the group $\Gamma=(N\ZZ)^d$ on 
$G$ by translations, i.e.\ $T_\gamma x= x - \gamma$ for all $\gamma 
\in \Gamma$ and $x \in G$. Define a colouring by $\cM\colon \ZZ^d \to 
C_N, \cM(x) \equiv x \mod \Gamma$.  Thus the colouring $\cM$ is 
$\Gamma$-periodic, i.e.~$\cM(x) = \cM(T_\gamma x)$ for all $\gamma 
\in\Gamma$ and $x\in G$.  In particular $\cM$ is uniquely defined by 
its values on the set $C_N\subset G$.  Let 
$A\colon\ell^2(G)\to\ell^2(G)$ be a selfadjoint finite-range operator 
which is $\cM$-invariant, in other words $\Gamma$-periodic.  This will 
be the deterministic ingredient determining the hopping part of the 
random operators we want to consider in this section. 
 
To define the random part, let $(\Omega,\PP)$ be a probability space 
and $\tau_\gamma\colon \Omega\to\Omega, \gamma\in\Gamma$, an ergodic 
family of measure preserving transformations.  Furthermore, let 
$\cala$ be an arbitrary finite subset of $\RR \cup \{+\infty\}$ and 
$(\omega,x)\mapsto V(\omega,x)\in \cala$ a random field which is 
invariant under the transformations $\tau_\gamma, \gamma\in \Gamma$. 
More precisely, for all $\gamma \in \Gamma$, $\omega \in \Omega$ and 
$x\in G$ we require $V(\tau_\gamma\omega,x)=V(\omega,T_\gamma x)$. 
Next we define random subsets of $G$ and $\ell^2(G)$ induced by the 
random field $V$.  For each $\omega\in \Omega$ define the subset of 
vertices $G_\omega:= \{x \in G : V(\omega,x) <\infty \}$, the natural 
projection operator $p_\omega\colon\ell^2(G)\to \ell^2(G_\omega)$ and 
its adjoint $i_\omega\colon\ell^2(G_\omega)\to \ell^2(G)$. 
 
The random Hamiltonian on $\ell^2(G_\omega)$ which we want to study is 
defined in the following way: the hopping part is given by 
\begin{equation*} 
  A_\omega := p_\omega A \, i_\omega, \quad \D(A_\omega):= \ell^2(G_\omega). 
\end{equation*} 
On the set $G_\omega$, the mapping $x \mapsto V(\omega,x) $ is a 
bounded, real-valued function and $V_\omega:=p_\omega V(\omega, \cdot) 
\, i_\omega \colon \ell^2(G_\omega)\to\ell^2(G_\omega)$ denotes the 
corresponding multiplication operator.  The total Hamiltonian is then 
given by 
\begin{equation*} 
  H_\omega := A_\omega + V_\omega, \quad \D(H_\omega):= \ell^2(G_\omega) 
\end{equation*} 
It is $\Gamma$-stationary in the sense that $U_\gamma H_\omega 
U_\gamma^*=H_{\tau_\gamma\omega}$, where $U_\gamma \phi(x):= \phi(x 
-\gamma)$ for all $x\in G$ and $\gamma\in\Gamma$.  Since $A_\omega$ is 
generated from the periodic operator $A$ by a site-percolation process 
on the graph $G$, it may be called a \emph{(site-) percolation 
  Hamiltonian}.  On the other hand, $V_\omega$ is a random potential 
as in the Anderson model on $\ell^2(\ZZ^d)$. For this reasons we call 
$H_\omega$, which contains features of both models, an 
Anderson-percolation Hamiltonian.  
For such operators the existence of the IDS as a pointwise limit 
has been established in \cite{Veselic-05a} and its continuity properties  
have been analysed in \cite{Ves} and a contribution to \cite{DodziukLPSV-06}. 
 
For each $\omega \in \Omega$ we define a colouring by 
\begin{equation*} 
  \gL_\omega\colon \ZZ^d \to \cala \times C_N, 
  \quad \gL_\omega(x):= (V(\omega,x),\cM(x)). 
\end{equation*} 
For any pattern $P \colon Q(P) \to \cala, Q(P)\in \cF(\ZZ^d)$, the 
frequency $\nu_P$ of $P$ in $\gL_\omega$ along the van Hove sequence 
of boxes $C_j,j \in\NN$, exists by the ergodic theorem and is 
independent of $\omega$ almost surely.  If $\nu_P$ is positive, let 
$\Omega_P$ consist of all $\omega\in \Omega$ for which the frequency 
of $P$ exists and equals $\nu_P$.  If $\nu_P$ is zero, then the set of $\omega'$ in $\Omega$ for which the pattern 
$P$ does occur at all has measure zero (as for each fixed position the 
set of $\omega'$ for which $P$ occurs at this position has measure 
zero and there are only countably many positions). In this case denote 
the complement of this set by by $\Omega_P$. In both cases, the set 
$\Omega_P$ has full measure.  Since there are only countably many 
finite subsets of $\ZZ^d$, the intersection $\Omega':=\bigcap_P 
\Omega_P$ has again measure one.  In the following we chose an $\omega 
\in \Omega'$ and keep it fixed.  \smallskip

To fit the operator $H_\omega$ in the abstract setting we have been 
considering in Sections 2 to 4, we extend it to the whole of 
$\ell^2(G)$ by setting it equal to zero on $\ell^2(G\setminus 
G_\omega)$. We denote this extension again by $H_\omega$. Thus the 
operator $H_\omega$ is $\gL_\omega$-invariant with 
$R_{\mathit{inv}}=1$ and is of finite range, since it is a sum of a 
finite-range and a diagonal operator. Therefore, Theorem \ref{ids} 
gives the existence of a unique probability measure $\mu_{H_\omega}$ 
on $\RR$ with distribution function $N_{H_\omega}$ such that 
$\frac{1}{|C_j|} n( p_{C_j} H i_{C_j})$ converges to $N_{H_\omega}$ 
with respect to the supremum norm: for every $j,M\in\NN$ we have the 
explicit estimate 
\begin{equation*} 
  \left \| \frac{n( p_{C_j} H i_{C_j}))  }{|C_j|} - N_{H_\omega} 
  \right\|_\infty   
  \leq  8 \frac{|\partial^R C_M|}{|C_M|}  +    
  \sum_{P\in\calpM}  \left| \frac{\sharp_P (\varLambda_\omega\cap 
      C_j)}{|C_j|} -  \nu_P\right| + 5 |C_R| \frac{|\partial^M C_j|}{|C_j|} . 
\end{equation*} 
The spectrum of $H_\omega$ and the topological support of 
$\mu_{H_\omega}$ coincide. Moreover, by Corollary \ref{huhu} and the 
fact that the frequencies are independent of $\omega\in \Omega'$, we 
see that the distribution function $N_{H_\omega}$ is in fact the same 
for all $\omega\in \Omega'$ and that the spectrum of 
$\sigma(H_\omega)$ as a set does not depend on $\omega\in \Omega'$. 
 
Let us close this section by pointing out to which situations the 
results 
presented here can be easily extended: \quad (1)\quad If the random 
part $V_\omega$ of the Hamiltonian  
$H_\omega$ is absent, we obtain a $\Gamma$-periodic, 
i.e.~$\cM$-invariant operator. This shows that in certain cases, one 
can use an alternative description of periodic operators by graph 
colourings to the one presented in Section \ref{s-periodic}.\\ 
(2)\quad On the other hand, one can combine the constructions from 
Section \ref{s-periodic} and the the present one to define random 
Hamiltonians on more general $\ZZ^d$-equivariant graphs than $\ZZ^d$ 
itself.\\ 
(3)\quad Analogous results to those derived here for 
\emph{site}-percolation Hamiltonians can be derived for Hamiltonians on 
\emph{bond}-percolation graphs by using the same arguments. For the 
construction of the IDS for such operators and for the asymptotics of 
the IDS at spectral edges, see \cite{KirschM-06, MuSt05}.

%%%%%%%%%%%%%%%%%%%%%%%%%%%%%%%%%%%%%%%%%%%%%%%%%%%%%%%%%%%%%%%%%%%%%%%%%%%%%%%%%%%%%%%%%%%%%%%%% 
 
\section{Application to visible points} 
 
The set of visible points in $\ZZ^d$ is a prominent example (and 
counterexample) in number theory and aperiodic order \cite{BMP,Ple}. 
In particular its diffraction theory has been well studied. Still, it 
seems that the corresponding nearest-neighbour hopping model has not 
received attention so far. In this section we provide a first modest 
step towards studying such a model by showing existence of the 
associated integrated density of states. 
 
The set $\calv$ of visible points in $\ZZ^d$ consists of the origin 
and all $x\neq 0$ in $\ZZ^d$ with 
\begin{equation*} 
  \{t x : 0< t <1\}\cap \ZZ^d =\emptyset. 
\end{equation*} 
Put differently, $x\neq 0$ belongs to $\calv$, if and only if the greatest common 
divisor of its coordinates is $1$. The obvious interpretation is that 
such an $x$ can be seen by an observer standing at the origin. This 
gives the name to this set. The characteristic function 
\begin{equation*}  
  \varLambda := \chi_\calv \colon \ZZ^d\longrightarrow \cala:=\{0,1\} 
\end{equation*}  
of $\calv$ provides a colouring. While $\calv$ is very regular in many 
respects, it has arbitrarily large holes.  In particular, existence of 
the frequencies $\nu_P$ does not hold along arbitrary van Hove 
sequences. However, as was shown in \cite{Ple} (see \cite{BMP} for special 
cases as well), the frequencies exist and can be calculated explicitly 
for sequences of cubes containing the origin.  Moreover, the 
frequencies of all patterns which occur are strictly positive. 
 
Thus, all abstract results discussed in this paper are valid for 
$\chi_\calv$-invariant operators of finite range. One relevant such 
operator is the hopping Laplacian $\Delta_\calv$. We finish this 
section by defining this operator: Points $x=(x_1,\ldots,x_d)$ and $y 
= (y_1,\ldots, y_d)$ in $\ZZ^d$ are said to be neighbours, written as 
$x\sim y$, whenever 
\begin{equation*}  
  \sum_{j=1}^d |x_j - y_j|=1. 
\end{equation*} 
Then, $\Delta_\calv \colon \ell^2 (\ZZ^d) \longrightarrow \ell^2 
(\ZZ^d)$ is defined by 
\begin{equation*} 
  (\Delta_\calv  u) (x) := \chi_\calv (x) \sum_{y\sim x: y\in \calv} 
  u(y) 
\end{equation*}  
for all $x\in\ZZ^{d}$ and all $u\in\ell^2 (\ZZ^d)$. 
Obviously, assumptions (S) and (+) are fulfilled for this operator, 
and the results of Section~\ref{sect:uniex} apply.

\section{Open questions} 
 
The considerations of the previous sections naturally raise various 
questions. Most prominent is the the question to what extent similar 
ergodic theorems hold on more general groups than $\ZZ^d$. More 
specifically, one may consider finitely generated groups which are  
amenable and/or residually finite. 
 
One may wonder about analogous statements for the IDS for suitable 
operators in continuous geometries.  As an intermediate step between 
continuum and discrete models one may consider quantum graphs.  In 
both cases the distributions in questions are no longer bounded. 
 
Finally, it may be interesting to learn more about the spectral theory 
of the nearest-neighbour hopping Laplacian on the set of visible points.

\subsection*{Acknowledgments}  
Part of this work was done while one of the authors (D.L.) was 
visiting G\"ottingen. He would like to thank Thomas Schick for the 
invitation.  He would also like to thank Michael Baake, Peter 
Pleasants and Bernd Sing for useful discussions concerning the set of 
visible points.  Partial support from DFG is gratefully acknowledged.

\bigskip~ 
\parskip0pt 
 
\end{document}